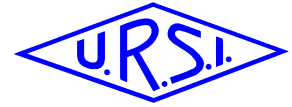

# *Fermi*, *Wind*, and *SOHO* Observations of Sustained Gamma-Ray Emission from the Sun

N. Gopalswamy[(1)], P. Mäkelä[(1,2)], S. Yashiro[(1,2)], A. Lara[(1,2,3)], H. Xie[(1,2)], S. Akiyama[(1,2)], R. J. MacDowall[(1)]
(1) NASA Goddard Space Flight Center, Greenbelt, Maryland, USA
(2) The Catholic University of America, Washington DC, USA
(3) Universidad Nacional Autónoma de México, CdMx, México
nat.gopalswamy@nasa.gov

## Abstract

We report on the linear relationship between the durations of two types of electromagnetic emissions associated with shocks driven by coronal mass ejections: sustained gamma-ray emission (SGRE) and interplanetary type II radio bursts. The relationship implies that shocks accelerate ~10 keV electrons (for type II bursts) and >300 MeV protons (for SGRE) roughly over the same duration. The SGRE events are from the Large Area Telescope (LAT) on board the Fermi satellite, while the type II bursts are from the Radio and Plasma Wave Experiment (WAVES) on board the Wind spacecraft. Here we consider five SGRE events that were not included in a previous study of events with longer duration ( >5 hours). The five events are selected by relaxing the minimum duration to 3 hours. We found that some SGRE events had a tail that seems to last until the end of the associated type II burst. We pay special attention to the 2011 June 2 SGRE event that did not have a large solar energetic particle event at Earth or at the STEREO spacecraft that was well connected to the eruption. We suggest that the preceding CME acted as a magnetic barrier that mirrored protons back to Sun.

## 1. Introduction

Sustained Gamma-Ray Emission (SGRE) from the Sun lasting for hours beyond the impulsive phase of the associated flares [1-2] are thought to be due to the decay of neutral pions produced when >300 MeV protons strike the solar chromosphere. While only a couple of such SGRE events were reported in the past [1-2], observations from the Large Area Telescope (LAT) on board the Fermi satellite [3] have shown that these events are rather common. Since its launch, Fermi has detected about 30 SGRE events until the end of 2017 [4-8]. Two possible sources of these protons have been discussed in the literature: protons accelerated/trapped in large-scale flare structures or by MHD shocks as reviewed in [9-10]. SGRE durations extending to almost a day clearly point to the shock source. Furthermore, even eruptions occurring on the backside of the Sun produce gamma-ray emission on the disk, indicating a particle source that extends over tens of degrees unlike flares [4, 11-13], analogous to the gamma-ray line emission from backside eruptions [14]. The shock possibility has been bolstered by the association between SGRE events and type II radio bursts in the decameter-hectometric (DH) wavelengths [5-6]. It is well known that the DH type II bursts are caused by interplanetary (IP) shocks driven by energetic coronal mass ejections (CMEs) characterized by high speeds and wide angular extent [15]. It was recently shown for the first time that Fermi/LAT >100 MeV SGRE durations are linearly related to the durations and ending frequencies of DH type II bursts [16] using 13 SGRE events with >5h duration. This finding strongly evidence that supports the idea that the same shock accelerates electrons causing type II bursts in the vicinity of the shock and protons producing SGREs after a sunward journey. This is consistent with the close association between type II bursts extending to kilometric wavelengths and large solar energetic particle (SEP) events [17].

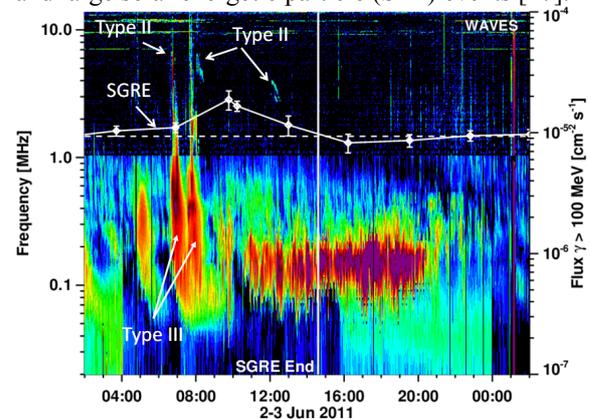

**Figure 1.** SGRE light curve plotted on a *Wind*/WAVES dynamic spectrum showing two fragments of a type II burst and a brief preceding type II burst. The two vertical features are type III bursts from two eruptions. The horizontal dashed line corresponds to the *Fermi*/LAT background. Three signal data bins can be seen above the background after 07:46 UT (soft X-ray flare peak). The SGRE end was dtermined according to our criterion.

There are some significant differences in the estimated durations of SGRE events among authors [5,7,16]. For example, the SGRE duration of the 2011 June 02 event was estimated in [5] as 3.83 hours, but only ~0.7 hours in [7]. The differences arise from the criteria used in defining the starting and ending times. Since both type II radio bursts and SGRE are detected above a background, a new criterion was established in [16] to determine the SGRE duration as follows. The start time was selected as the peak time of the associated soft X-ray flare to avoid the impulsive phase gamma rays. The end time was defined as the mid time between the last signal data point (including the error bar) above the background and the next data point. According to this criterion, the 2011 June 02 SGRE has a duration of 6.84±1.61 hours. The error bar is half of the time interval between the last two data points. Fig. 1 shows this SGRE event along with the type II radio burst from Wind/WAVES. Clearly, our end time is after the last signal data point above the background, but before the next data point and yields SGRE flux still above the background. The estimated duration of 3.83 h in [5] corresponds to the last signal data point, but that in [7] is clearly an underestimate.

In this paper, we consider five additional SGRE events to the list published in [16] applying the new criterion mentioned above. The primary list of events is still from [5], but the durations are estimated using our criterion. It is shown that the new events are consistent with the linear relationship established between the SGRE duration and the type II duration. In addition, we discuss some peculiarities in these events that provide a better understanding of the energetic protons involved in the SGRE events.

**Table 1.** SGRE, CME, flare, and type II information

| SGRE | | CME | | Flare | SEP | Type II | |
|---|---|---|---|---|---|---|---|
| Start UT | Dur hr | UT | V km/s | Cls | >10 MeV | Start UT | Dur Hr |
| 11/6/2 07:46 | 6.84 ±1.61 | 0812 | 1147 H | C3.7 | 0.10 | 0800 | 4.47 ±0.02 |
| 11/6/7 06:41 | 9.47 ±1.35 | 0649 | 1321 H | M2.5 | 72 | 0645 | 10.93 ±0.32 |
| 12/1/27 18:37 | 9.52 ±0.73 | 1827 | 2541 H | X1.7 | 795 | 1830 | 10.23 ±0.33 |
| 12/5/17 01:47 | 3.09 ±0.80 | 0148 | 1596 H | M5.1 | 255 | 0140 | 4.73 ±0.20 |
| 13/5/13 02:17 | 7.47 ±1.62 | 0200 | 1270 H | X1.7 | 25 | 0220 | 6.12 ±0.28 |

## 2. II. SGRE and Type II Burst Durations

Table 1 shows the list of new events that are analyzed with the soft X-ray flare time taken as the SGRE start time (column 1). The SGRE durations (column 2) are determined using our criterion described in section I. The first appearance times and cone-model speeds of the associated CMEs are listed in columns 3 and 4. The flare class and SEP intensity (particle flux unit, pfu) are in columns 5 and 6. The DH type II onset time and durations are given in columns 7 and 8. The source locations of the CMEs are: 2011 June 2: S19E25, 2011 June 7: S21W54, 2012 January 27: N27W71, 2012 May 17: N11W76, and 2013 May 13: N11E90. The durations range from 3.09 to 9.52 h and correspond to the lower end considered in the previous paper [16]. The SGRE and type II durations are similar in all cases. In two cases (2011 June 7 and 2012 January 27), there was a tail of SGRE with at least one point above the background. The listed durations in Table 1 include the tails. Without the tails, the SGRE durations are 3.07±1.67 hours and 3.60±0.83 hours, respectively. In both cases, there was clear type II emission until the end of the tail. Figure 2 shows the SGRE light curve plotted on the Wind/WAVES dynamic spectrum. Note that the type II burst is intense and ends abruptly around 04:45 UT on January 28. The type II burst is not seen well at frequencies above 1 MHz mainly because the sensitivity of the high-frequency receiver is poorer than that of the low-frequency receiver. Including the tail in determining the SGRE duration matches well with the type II burst duration. It is interesting that the presence of the tail makes the overall SGRE duration consistent with the type II burst duration. It is worth noting that type II bursts require ~10 keV electrons somewhere on the shock front, whereas the gamma rays require >300 MeV protons near the nose region of the shock. It must be noted that the SGRE tail would not be normally identified as an event because it is too close to the background and there are data points below the average background determined over 4 days.

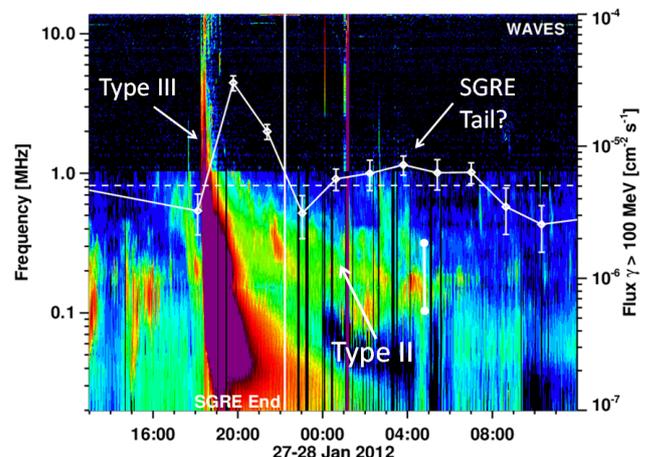

**Figure 2**. Dynamic spectrum and SGRE light curve of the 2012 January 27 event. The end of the type II burst is marked by the short vertical line. The long vertical line indicates the SGRE end (22:13 UT) following the criterion. Including the tail, the SGRE end time become 04:32 UT on January 28, giving the duration as 9.52 ±0.73 hours.

Figure 3 shows how the SGRE duration is related to the duration and ending frequency of type II bursts. The regression line was obtained using 13 SGRE events that had durations exceeding ~5 hours. The 1991 June 11 event [2] from the Energetic Gamma Ray Experiment Telescope (EGRET) with the associated type II from the Ulysses Unified Radio and Plasma (URAP) wave receiver and the events in Table 1 are in good agreement with the best-fit line obtained before and all the new data points are within the 95% confidence interval obtained for the 13 events. Thus the linear relationship holds for all 18 SGRE events with durations >3 hours. The relationship indicates that the shocks accelerating ~10 keV electrons also accelerate >300 MeV protons over roughly the same interval since its formation near the Sun driven by energetic CMEs. The anti-correlation between SGRE duration and type II ending frequency indicates that shocks accelerate protons over larger heliocentric distance in longer-duration SGRE events. Lower-frequency type II emission comes from larger heliocentric distances where the local plasma frequency is lower.

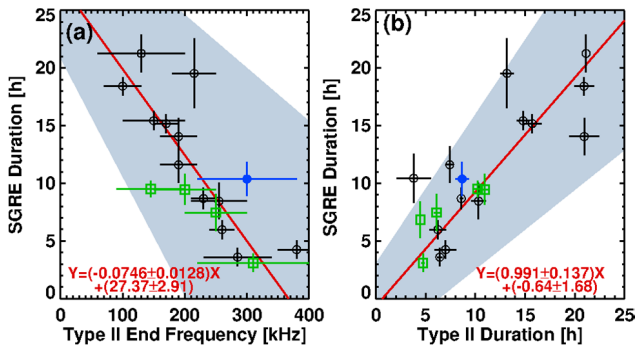

**Figure 3.** SGRE durations plotted against the type-II ending frequency (a) and type II duration (b) for 13 events with duration exceeding ~5 hours extracted from [5], but the durations were determined using the criteria mentioned in the previous section. The red lines are the best-fit lines obtained using the Orthogonal Distance Regression method. The equations of the best-fit lines are shown on the plots. The shaded area represents 95% confidence interval of the fit. The blue (1991 June 11 EGRET event) and green (events in Table 1) points are excluded in the fit.

## 3. Energetic Protons

Table 1 shows that the three well connected events, viz., 2011 June 7 (S21W54), 2012 January 27 (N27W71), 2012 May 17 (N11W76) are associated with large solar energetic particle (SEP) events. The >10 MeV proton intensities listed in Table 1 are in pfu. These events had intense flux of >100 MeV protons, indicating the likely existence of >300 MeV protons needed for SGREs. One of the well-connected events (the 2012 May 17) had ground level enhancement (GLE) indicating GeV particles reaching Earth, so the presence of >300 MeV protons is expected. The 2013 May 13 event was from the east limb, so there was no SEP event at Earth. However, there was a large SEP event detected by *STEREO-Behind* (*STB*) spacecraft (>10 MeV intensity was ~25 pfu as listed in Table1). The 2011 June 2 event did not have an SEP event at Earth. The source location was eastern (S19E25). *STB* was located at E93 at the time of this event, so the eruption was well connected to *STB* (S19W68 in *STB* view). Surprisingly, there was no large SEP event at *STB* either. There was only a slight increase in >10 MeV protons from a background of 0.02 pfu to a maximum of 0.1 pfu. The lack of a large SEP event is certainly puzzling and seemingly contradicts the idea of shock-accelerated protons causing the gamma-ray emission. The flare idea does not work because the flare was weak (C3.7) starting, peaking, and ending at 07:22, 07:46, and 07:57 UT, respectively. The SGRE lasted for about 7 hours after the end of the flare (see Fig. 4). The presence of shock itself is unquestionable because of the DH type II associated with a fast (1147 km/s) halo CME (Table 1). Although fragmented, the overall Type II duration roughly matches that of SGRE.

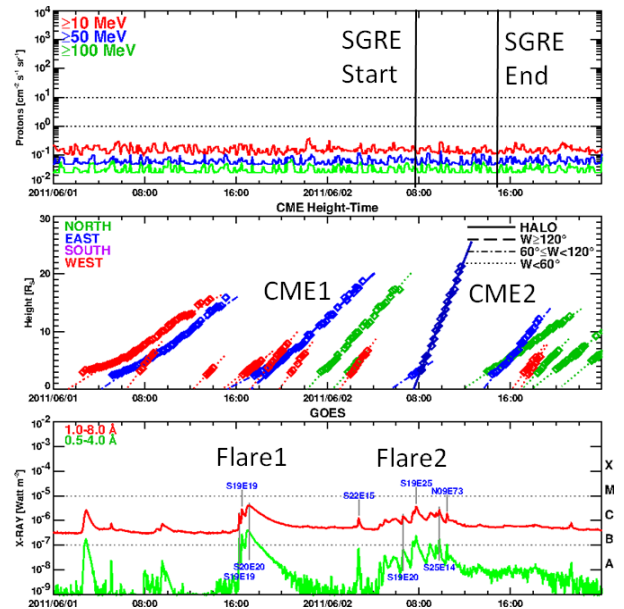

**Figure 4.** Proton intensity (top), CME height-time plots (middle) and *GOES* soft X-ray flares (bottom) around the 2011 June 2 event. CME2 (08:12 UT, June2) is associated with the SGRE and Flare2 at 07:22 UT. The preceding CME1 (18:36 UT, June 1) is associated with Flare1. Extrapolation of the height-time plot reveals that CME2 caught up with CME1 at ~65 Rs, just after the SGRE end.

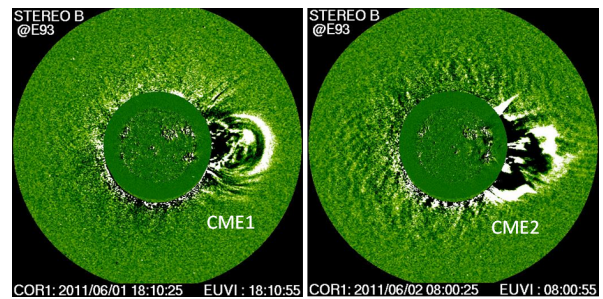

**Figure 5.** CME1 and CME2 as observed by *STEREO*'s inner coronagraph COR1. The CMEs had excellent position angle and longitudinal overlap. The CMEs were also separated by ~10 hours in the onset, so CME2 remained behind CME1 over the duration of the SGRE and caught up at a distance of ~65 Rs.

We suggest that the presence of the preceding CME (CME1) that first appeared in *SOHO* field of view at 18:36 UT on June 1 might explain the lack of SEP event at *STB*. In Fig. 5 this CME is marked as CME1 associated with a C2.9 flare (Flare1) from active region (AR) 11226 (S20E20). The CME that was responsible for the SGRE occurred about 10 hours later (08:12 UT on June 02) from the neighboring AR 11227 (S19E25) and associated with the C3.7 flare (Flare2). The projected speed of CME1 was 361 km/s, but likely to be much higher because of the disk-center (S20E20) location. CME1 was also partial halo. A plausible scenario is that CME1 blocked the outward propagation of protons from CME2, but the protons were able to propagate toward the Sun to produce the gamma rays. This is similar to the scenario often invoked in GLE events [18]: a structure causes a localized constriction of the magnetic field lines leading to mirroring of particles toward the Sun. In fact, some of the SGRE events with weak type II bursts have such preceding CMEs, indicating that CME interaction is an additional factor that influences the production of SGREs. Figure 5 shows the two west-limb CMEs in *STB* view from the inner coronagraph (COR1 (the source regions were located just to the east of the disk center in *SOHO* view). There was another CME with a brief DH type II immediately preceding CME2 might have contributed seed particles (see Fig. 1).

## 4. Summary and Conclusions

We investigated five SGRE events and identified the associated CMEs, type II radio bursts, soft X-ray flares, and SEP events. In all five cases, there were fast CMEs that were full halos. The speeds were in the range 1147 km/s to 2541 km/s. The average speed (1575±564 km/s) is slightly smaller than that (2094±621 km/s) of the longer duration events reported earlier [16]. There seems to be a tendency for longer-duration SGREs to be associated with faster CMEs. The high CME speeds are consistent with those producing DH type II bursts and large SEP events. The DH type II bursts indicate that the associated shocks propagated far into the interplanetary medium. The close relationship between SGRE and type II burst durations provides the strongest evidence that the same shock is responsible for accelerating the underlying particles (electrons for type II bursts and protons for SGRE).

All the well-connected events were associated with large SEP events and intense >100 MeV proton events. One of the two eastern events had a large SEP event at *STB*. The other eastern event (2011 June 2) was not associated with a large SEP event. The associated flare was the weakest (C3.7) of the 30 SGRE events reported in [5]. The evidence for the shock comes from the DH type II burst and fast CME. Since the overall duration of the SGRE and type II durations were similar, we surmise that protons were accelerated but did not reach the well-connected observer. A plausible scenario is that the magnetic structure of the preceding CME mirrored the protons from the primary CME toward the Sun causing SGRE.

## 5. Acknowledgements

We thank G. H. Share, B. R. Dennis, E. W. Cliver, J. M. Ryan, A. K. Tolbert and N. Omodei for helpful discussion. We benefited from the open data policy of *Fermi*, *SOHO*, *STEREO*, *Ulysses*, and *Wind* teams. Work supported by NASA's LWS and Heliophysics GI programs.

## 7. References

1. V. V. Akimov, V. G. Afanassyev, A. S. Belaousev et al., "Observation of High Energy Gamma-rays from the Sun with the GAMMA-1 Telescope (E > 30 MeV)", Proc. 22nd ICRC, vol. 3, 73, 1991.
2. G. Kanbach, D. L. Bertsch, C. E. Fichtel, et al., "Detection of a long-duration solar gamma-ray flare on June 11, 1991 with EGRET on COMPTON-GRO", Astron. Astrophys. Sup., vol. 97, 349, 1993.
3. M. Ajello, A. Albert, A Allafort, et al., "Impulsive and Long Duration High-energy Gamma-Ray Emission from the Very Bright 2012 March 7 Solar Flares", Astrophys. J., vol. 789, 20, 2014.
4. M. Ackermann, A. Allafort, L. Baldini et al., "Fermi-LAT Observations of High-energy Behind-the-limb Solar Flares", Astrophys. J., vol. 835, 219, 2017
5. G. H. Share, R. J. Murphy, A. K. Tolbert et al., "Characteristics of Sustained >100 MeV Gamma-ray Emission Associated with Solar Flares", Astrophys. J. Suppl. S. submitted, 2017arXiv171101511S.
6. K.-L. Klein, K Tziotziou, P. Zucca et al., "X-Ray, Radio and SEP Observations of Relativistic Gamma-Ray Events", Astrophys. Space Sci. L., vol. 444, 133, 2018.
7. L. M. Winter, V. Bernstein, N. Omodei et al., "A Statistical Study to Determine the Origin of Long-duration Gamma-Ray Flares", Astrophys. J., vol. 864, 39, 2018.
8. N. Omodei M. Pesce-Rollins, F. Longo, et al., "Fermi-LAT Observations of the 2017 September 10 Solar Flare", Astrophys. J., vol. 865, L7, 2018.
9. E. L. Chupp and J. M. Ryan "High energy neutron and pion-decay gamma-ray emissions from solar flares", Res. Astron. Astrophys., vol. 9, 11, 2009.
10. N. Vilmer, A. L. MacKinnon, Hurford, G., "Properties of Energetic Ions in the Solar Atmosphere from γ-Ray and Neutron Observations", Space Sci. Rev., vol. 159, 167, 2011.
11. M. Pesce-Rollins, N. Omodei, V. Petrosian et al., "First Detection of >100 MeV Gamma Rays Associated with a Behind-the-limb Solar Flare", Astrophys. J., vol. 805, L15, 2015.
12. Plotnikov, I., Rouillard, A. P., and Share, G. H., "The magnetic connectivity of coronal shocks from behind-the-limb flares to the visible solar surface during gamma-ray events", Astron. Astrophys., vol. 608, A43, 2017.
13. M. Jin, V. Pertosian, N. V. Nitta et al., "Probing the Puzzle of Behind-the-Limb Gamma-Ray Flares: Data-driven Simulations of Magnetic Connectivity and CME-driven Shock Evolution", Astrophys. J. accepted, 2018arXiv180701427J, 2018.
14. E. W. Cliver, D. J. Forrest, H. V. Cane et al., "Solar flare nuclear gamma-rays and interplanetary proton events", Astrophys. J., vol. 343, 953, 1989.
15. N. Gopalswamy, S. Yashiro, M. L. Kaiser et al., "Characteristics of coronal mass ejections associated with long-wavelength type II radio bursts", J. Geophys. Res., vol. 106, 29219, 2001.
16. N. Gopalswamy, P. Mäkelä, S. Yashiro et al., "Interplanetary Type II Radio Bursts from Wind/WAVES and Sustained Gamma Ray Emission from Fermi/LAT: Evidence for Shock Source", Astrophys. J. Lett. Under review, 2018.
17. N. Gopalswamy, "Coronal Mass Ejections and Type II Radio Bursts", Geoph. Monog. Series, Ed. Gopalswamy, N., Mewaldt, R., and Torsti, J., vol. 165, 207, 2006.
18. J. W. Bieber, W. Dröge, P. A. Evenson, et al., "Energetic Particle Observations during the 2000 July 14 Solar Event", Astrophys. J., vol. 567, 622, 2002